\documentstyle[aps,pra,epsfig,floats]{revtex} 

\makeatletter
\newcommand{\ps@inprint}{%
\renewcommand{\@oddhead}{}%
\renewcommand{\@evenhead}{}%
\renewcommand{\@oddfoot}{%
\begin{picture}(0,0)%
\put(-30,-20){\rotatebox{90}{\hspace*{200pt}%
\tt\underline{to appear in Physical Review A}}}%
\end{picture}%
\hfil\thepage\hfil}
\renewcommand{\@evenfoot}{\@oddfoot}}

\newlength{\septumdist}
\newcommand{\septum}{\par\vspace*{0.5\septumdist}\par%
\noindent\hspace*{\fill}\rule[-2pt]{1in}{0.5pt}\hspace*{\fill}%
\par\vspace*{0.5\septumdist}\par}
\def\references{%
\ifpreprintsty
\newpage
\hbox to\hsize{\hss\large \refname\hss}%
\else\septum\fi
\list{\@biblabel{\arabic{enumiv}}}%
{\labelwidth\WidestRefLabelThusFar  \labelsep4pt %
\leftmargin\labelwidth %
\advance\leftmargin\labelsep %
\ifdim\baselinestretch pt>1 pt %
\parsep  4pt\relax %
\else %
\parsep  0pt\relax %
\fi
\itemsep0pt plus 2pt\relax%
\usecounter{enumiv}%
\let\p@enumiv\@empty
\def\theenumiv{\arabic{enumiv}}%
}%
\let\newblock\relax %
\sloppy\clubpenalty4000\widowpenalty4000
\sfcode`\.=1000\relax
\ifpreprintsty\else\small\fi
}
\def\endreferences{%
\def\@noitemerr{\@warning{Empty `thebibliography' environment}}%
\endlist     \let\@SetMaxRefLabel\@gobble
}
\makeatother
\begin{document}
\draft
\title{Quantitative~wave-particle~duality
and~non-erasing~quantum~erasure}
\author{Peter D. D. Schwindt,$^{\dagger}$
Paul G. Kwiat,$^{\dagger}$ and
Berthold-Georg Englert$^{\ddagger,\ast}$}
\address{$^{\dagger}$Physics Division, P-23,
Los Alamos National Laboratory, Los Alamos, NM 87545\\
$^{\ddagger}$Max-Planck-Institut f\"{u}r Quantenoptik,
Hans-Kopfermann-Str.~1, 85748 Garching, Germany\\
$^{\ast}$%
Department of Physics, Texas A\&M University,
College Station, TX 77843-4242}
\date{Received 3 May 1999}
\wideabs{%
\maketitle
\begin{abstract}\hspace*{1em}%
The notion of wave-particle duality may be quantified by the inequality
$V^2+K^2\le1$, relating interference fringe visibility $V$ and path
knowledge $K$.
With a single-photon interferometer in which polarization is used to 
label the paths, we have investigated the relation for various 
situations, including pure, mixed, and partially-mixed input states.
A quantum eraser scheme has been realized that recovers interference 
fringes even when no which-way information is available to erase.
\end{abstract}
}

\thispagestyle{inprint}
\enlargethispage{0.8\baselineskip}

Wave-particle duality (WPD) dates back to Einstein's seminal paper on
the photoelectric effect \cite{ref:Einst}, and is a striking manifestation
of Bohr's complementarity principle \cite{ref:Bohr} (for a formal
definition see Ref.~\cite{ref:SEWnat}).
The familiar phrase ``each experiment must be described either in terms
of particles or in terms of waves'' emphasizes the extreme cases and
disregards intermediate situations, in which particle and wave aspects
coexist.
Theoretical investigations \cite{ref:WZ,ref:WPDbge1}, supplemented by a few 
experimental studies \cite{ref:WPDexp1,ref:Rempe1}, have led to a
quantitative formulation of WPD [Eq.\ (\ref{eq:WPDineq}) below].
Here we report an experiment using a single-photon Mach-Zehnder 
interferometer in which polarization marks the path.
We investigated the entire scope of the duality relation, for pure, 
mixed, and partially-mixed input states, and found {\em absolute\/} 
agreement at the percent level \cite{ref:Rempe2}.
We also realized a novel quantum eraser scheme, whereby interference is 
recoverable although no which way (WW) information was available to 
erase.
In view of the kinematical equivalence of all binary degrees of freedom,
our results are directly applicable whenever an interfering particle is 
entangled with a 2-state quantum system.

To quantify WPD, one needs quantitative, measurable characteristics for 
the wave-like and particle-like behavior of quanta.
In an interferometer, the former is naturally quantified by the 
{\em Visibility\/} $V$ of the observed interference fringes.
The quantification of the latter is based on the {\em Likelihood\/}
$L$ of correctly guessing the path taken by a particular quantum
--- the better one can guess, the more pronounced are the particle
aspects.
A random guess gives $L=\frac{1}{2}$, whereas $L=1$ indicates that the 
way is known with certainty.
The actual WW {\em Knowledge\/} $K$ is given by $K=2L-1$, with 
${0\leq K\leq1}$.
In an {\em asymmetric\/} interferometer one way is more likely
than the other to begin with (${L_{\rm a~priori}>\frac{1}{2}}$);
we call WW knowledge of this kind {\em Predictability\/}
(${P=2L_{\rm a~priori}-1}$).
The statement $V^2+P^2\leq1$ has been known for some time, implicitly
or explicitly, in various physical contexts \cite{ref:WZ,ref:WPDexp1}.
Since one cannot lose {\em a priori\/} knowledge, ${P\leq K}$;
in fact, $P\approx 0$ in our experiments.
Nevertheless, owing to an {\em entanglement\/} of the system wave
function with the wave function of some WW marker (WWM), the
Knowledge can still be as large as $1$.

The actual value of $K$ depends on the ``betting strategy''
employed; the optimal strategy maximizes K and identifies the
{\em Distinguishability\/} $D=\max\{K\}$ --- it
is the maximum amount of WW Knowledge available, although
a non-optimal measurement may yield less, or even zero.
(Experimental inaccessibility of some crucial degrees of freedom
may force the experimenter to settle for a non-optimal measurement;
see Ref.~\cite{ref:BjKa} for further remarks.)
Except where noted, our measurements were suitably optimized to
maximize $K$.
The {\em Duality Relation\/} accessible to experimental test then
becomes \cite{ref:WPDbge1,ref:BjKa}:
\begin{equation}\label{eq:WPDineq}
V^2 + K^2 \leq 1 \;.
\end{equation}
The equality holds for pure initial states of the WWM,
while the inequality applies to (partially-)mixed states.

\enlargethispage{0.8\baselineskip}

\section{Experimental setup and procedure}\label{sec:setup}
\begin{figure}[!t]
\noindent\begin{picture}(246,152)
\put(-2,10){\psfig{figure=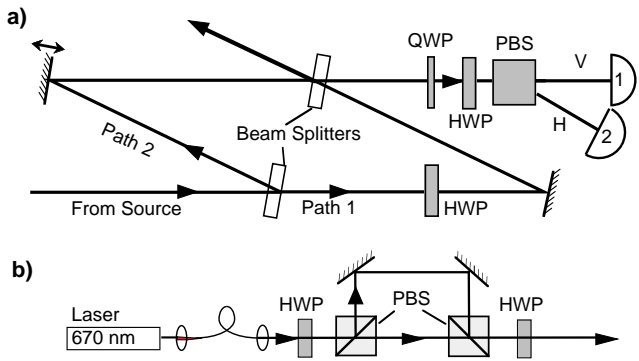,width=\columnwidth}}
\end{picture}
\caption[setup]{\label{fig:setup}
(a)~Experiment to quantify the wave-particle duality.
(b)~An asymmetric Mach-Zehnder interferometer
with polarizing beam splitters separates the horizontal and
\mbox{vertical}
components of the input light by much more than the
$\sim 1$-cm coherence length;
one can thus make arbitrary quantum states
(pure, mixed, and partially-mixed)
by varying the input polarization.
A single-mode fiber filters the spatial mode.}
\end{figure}

In our experiments single photons (at $670\,{\rm nm}$) were directed 
into a compressed Mach-Zehnder interferometer \cite{ref:compressed}
(see Fig.~\ref{fig:setup}).
An adjustable half waveplate (HWP) in path 1 was used to entangle
the photon's path with its polarization (i.e., with the WWM)
thus yielding WW Knowledge \cite{ref:waveplate}.
Our adjustable analysis system --- quarter waveplate (QWP), HWP, and
calcite prism (PBS) --- allowed the polarization WWM to be measured 
in any arbitrary basis.
The photons were detected using geiger-mode avalanche photodiodes ---
Single Photon Counting Modules (EG\&G \#SPCM-AQ, efficiency $\sim60\%$).
The input source, described below, was greatly attenuated so that the
maximum detection rates were always less than $50,000{\rm s}^{-1}$;
for the interferometer passage time of 1ns, this means that on average
fewer than $10^{-4}$ photons were in the interferometer at any time.

The probability for having no photon at all is close to unity at any
arbitrary instant, but state reduction removes this part a posteriori 
as soon as a detector ``clicks.''
The reduced state is virtually indistinguishable from a one-photon 
Fock state because the probability for two or more photons is 
negligibly small.
This one-photon-at-a-time operation is essential to allow sensible
discussion of the likely path taken by an individual light
quantum~\cite{ref:QKrypt}.

Perhaps unnecessarily, we emphasize that our experiment is not intended 
to be a direct proof of the quantum  nature of light.
Rather, we  accept the existence of photons as an established 
experimental fact \cite{ref:photon1}.
The quantized electromagnetic field has a classical limit as a field 
(unlike other quantum fields that have, at best, a limit in terms of
particles), and some properties of the quantum field have close 
classical analogs.
In particular, the counting rates of single-photon interferometers, 
such as the one used in our experiment, are proportional to the 
intensities of the corresponding classical electromagnetic field. 
But there is no allowance for individual detector clicks in Maxwell's
equations \cite{ref:clicks}, nor for the quantum entanglement of 
photonic degrees of freedom that we exploit.
And clearly, the trajectory of a light quantum through the interferometer 
is a concept alien to classical electrodynamics, as is the experimenter's
knowledge $K$ about this trajectory.

For Visibility measurements the polarization analyzer was lowered
out of the beam path, and the maximum and minimum count rates
on detector 1 were measured as the length of path 2 was varied
slightly (via a piezoelectric transducer).
After subtracting out the separately-measured detector background 
(i.e., the count rate when the input to the interferometer was blocked, 
typically 100--400${{\rm s}^{-1}}$), the visibility was calculated in 
the standard manner: $V=({\rm Max}-{\rm Min})/({\rm Max}+{\rm Min})$.

For the determination of the Likelihood, and hence the Knowledge,
the following procedure was used.
With the polarization analyzer in place, and path 2 blocked, the
counts on the two detectors were measured.
Detector 1 (2) looked at polarization $\lambda$ ($\lambda^{\perp}$),
determined by the analysis settings.
After subtracting the backgrounds measured for each detector, the count
rates from detector 1 were scaled by the relative efficiency of the two
detectors: $\eta_2/\eta_1{=1.11\pm0.01}$.
(In this way our calculated value of the Knowledge corresponds to what
would have been measured if our detectors had been identical and
noiseless.)
Call the resulting scaled rates
$R_{1\lambda}\equiv R(\mbox{path 1}, \mbox{polarization $\lambda$})$
and $R_{1\lambda^{\perp}}\equiv %
R(\mbox{path 1}, \mbox{polarization $\lambda^{\perp}$})$.
Next, we measure the corresponding quantities for path 2:
$R_{2\lambda}$ and $R_{2\lambda^{\perp}}$.
The betting strategy is the one introduced by Wootters and Zurek
\cite{ref:WZ} and optimized in Ref.~\cite{ref:WPDbge1}:
Pick the path which contributes most to the probability of
triggering the detector that has actually fired.
The Likelihood is then 
\begin{equation}
L = \frac {\max\{R_{1\lambda},R_{2\lambda}\}
+ \max\{R_{1\lambda^{\perp}},R_{2\lambda^{\perp}}\}}
{R_{1\lambda} + R_{2\lambda} +
 R_{1\lambda^{\perp}} + R_{2\lambda^{\perp}}} \,.
\end{equation}

\section{Experimental results}\label{sec:results}
\subsection{Wave-particle duality for pure states}\label{ssec:WPDpure}
\begin{figure}[!t]
\noindent\begin{picture}(246,137)
\put(-2,10){\psfig{figure=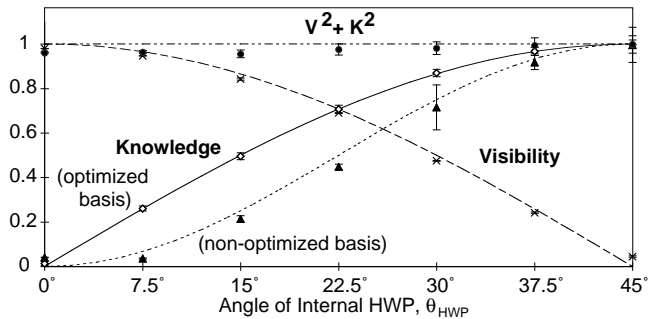,width=\columnwidth}}
\end{picture}
\caption[pureVK]{\label{fig:pureVK}
Experimental data and theoretical curves
for a purely vertically-polarized input to the system shown in
Fig.~\ref{fig:setup}, as a function of the orientation
of a HWP in path 1.
The crosses are the measured Visibilities;
the dotted line and triangles correspond to $K$ measurements fixed in
the horizontal-vertical basis; and the solid line and diamonds
correspond to measurements in the optimal basis \cite{ref:optimal}.}
\end{figure}

Figure \ref{fig:pureVK} shows the results when a pure
vertical-polarization state ({\sf V}) was input to the interferometer,
as a function of the internal HWP's orientation.
As expected, when the HWP is aligned to the vertical
($\theta_{\rm{HWP}} = 0$), therefore leaving the polarization
unchanged, we see nearly complete Visibility and get no WW Knowledge.
The measured values of $V$ are slightly lower than
the theoretical curve because the intrinsic visibility of the
interferometer (even without  the HWP) is only
$\sim 98\%$, due to nonideal optics \cite{ref:visibility}.
Conversely, with the HWP set (at $\theta_{\rm{HWP}} = 45^{\circ}$)
to rotate the polarization in path 1 to horizontal ({\sf H}), the
Visibility is essentially zero, and the Knowledge nearly equal to 1.
Formally, the spatial wave function and the polarization WWM wave
function are completely entangled by the HWP:
${|\psi\rangle \propto |1\rangle|{\sf H}\rangle_{{\rm WWM}}
+ e^{i \phi}|2\rangle|{\sf V}\rangle_{{\rm WWM}}}$,
where $\phi$ is the relative phase between paths 1 and 2.
Tracing over the WWM effectively removes the coherence
between the spatial modes.
That a small visibility persists in our results can be explained
by slight residual polarization transformations by the interferometer
mirrors and beam splitters, so that the polarizations from the two
paths are not completely orthogonal;
and by the remarkable robustness of interference --- both theoretically
and experimentally, $V>4.4\%$ even though $L>99.9\%$!

In Fig.\ \ref{fig:pureVK} we also display two sets of Knowledge data,
one taken in the optimal basis \cite{ref:optimal}, the other fixed in the 
horizontal-vertical basis.
These data demonstrate that Knowledge can depend on the measurement
technique.
With the optimal basis, the value of ${V^2 + K^2}$ is always very close 
to the predicted unit value; our experiment is the first to verify this.
The average of all the data points in Fig.~\ref{fig:pureVK} gives
${0.976\pm0.017}$.
The slight discrepancy with the predicted value of $1$ is mostly due to
the intrinsic visibility of the interferometer --- for the
minimum-visibility arrangement, ${V^2+K^2=0.998}$.

\subsection{Wave-particle duality for (partially-)mixed states}%
\begin{figure}[!t]
\noindent\begin{picture}(246,275)
\put(-2,10){\psfig{figure=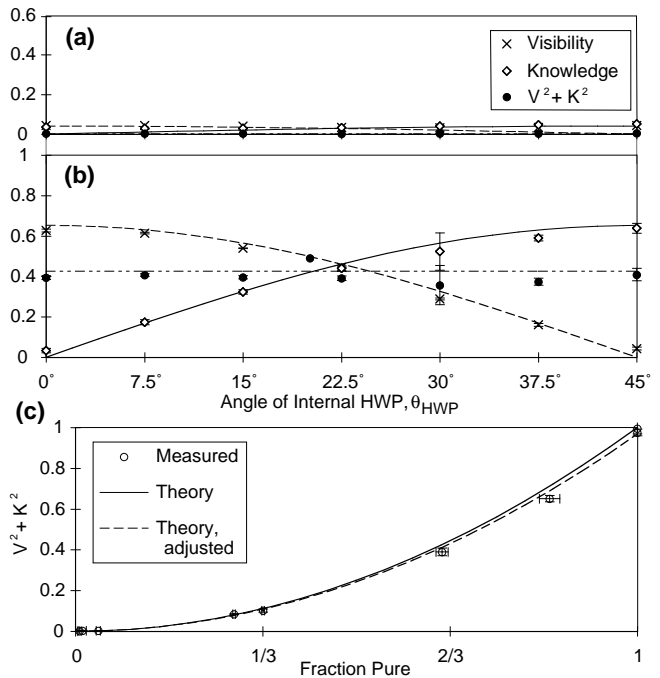,width=\columnwidth}}
\end{picture}
\caption[mixedVK]{\label{fig:mixedVK}
Visibility and Knowledge measurements (and theory curves) for various
mixed and partially-mixed input states.
(a) A mixed state input  from the filtered white-light source yielded
an average value for $V^2 + K^2$ of $0.003\pm0.001$;
the theoretical prediction based on the measured input state is
$0.017\pm0.003$ (the uncertainty in the theory comes from
imperfect determination of the state).
The slight disparity arises from residual polarization transformations
by the empty interferometer.
(b) A partially-mixed state (purity = $0.65 \pm 0.01$)
was generated using the tunable source.
(c) A summary of all $V^2 + K^2$ data for various input states.
The solid curve is the uncorrected theory:
$V^2+D^2=2{\rm Tr}(\rho^2)-1$, where $\rho$ is the density matrix of
the polarization WWM. 
The dashed curve is the theory accounting for the maximum visibility
and the slight polarization-dependence of the empty interferometer.}
\end{figure}

\label{ssec:WPDmix}
Using photons from an attenuated quartz halogen lamp that was
spectrally-filtered with a narrow-band interference filter (centered
at $670\,{\rm nm}$, $1.5\,{\rm nm}$ FWHM) and spatially-filtered
via a single-mode optical fiber, we explored Eq.\ (\ref{eq:WPDineq}) for 
{\it mixed} states (slight polarizing effects from the fiber actually 
led to $\sim4\%$ residual net polarization).
The measurements of Visibility and Knowledge for this nearly 
completely-mixed input state have values close to the theoretical 
prediction of $0$ (Fig.\ \ref{fig:mixedVK}a).
$K \rightarrow 0$ for a completely-mixed WWM state because
any unitary transformations on an unpolarized input also
yield an unpolarized state (the density matrix
is unaffected), so there is no WW information.
That $V \rightarrow 0$ can be understood by examining the behavior of
orthogonal pure WWM states, with no definite phase relationship between 
them.
In the basis where the HWP rotates the WWM states by $90^{\circ}$,
the orthogonal polarizations from paths 1 and 2 cannot interfere;
in the basis aligned with the HWP's axes, each polarization
individually interferes, but the interference patterns are
shifted relatively by $180^{\circ}$ (due to the birefringence of
the HWP), so the sum is a fringeless constant.

To enable production of an even more mixed input, and
to allow generation of arbitrary partially-mixed states, we used
a ``tunable'' diode-laser scheme (see Fig.\ \ref{fig:setup}b).
By rotating the (pure linear) polarization input to the first
polarizing beam splitter, one can control the relative contribution
of horizontal and vertical components.
For example, for incident photons at $45^{\circ}$, one has equal
{\sf H} and {\sf V} amplitudes which are then added together with a
random and rapidly varying phase to produce an effectively completely
mixed state of polarization \cite{ref:LLP}.
With 5 times more vertical than horizontal, the state is then
1/3 completely-mixed to 2/3 pure.
This case is shown in Fig.~\ref{fig:mixedVK}b.
Note that the maximum Visibility (and Knowledge) is numerically equal
to the state purity, as the mixed-component displays no
interference (and contains no WW information).
The data taken for various input states show excellent agreement with 
theoretical predictions (Fig. \ref{fig:mixedVK}c).

\subsection{Quantum erasure (erasing and non-erasing)}\label{ssec:QE}
In contrast to many interference situations where the WW information 
may be inaccessible, the quantum state of our WWM is easily manipulated.
One can then in fact ``erase'' the distinguishing information and
recover interference \cite{ref:SEWnat,ref:Scully2} (though this simple
physical picture fails when non-pure WWM states are considered).
In our experiments, such an erasure consists of using a polarization
analysis to reduce or remove the WW labeling. 
For example, if the path 1 and 2 polarizations are horizontal
and vertical, respectively, analysis at $\pm45^{\circ}$ will recover
complete fringes; any photon transmitted through a $45^{\circ}$ 
polarizer is equally likely to have come from either path.

\begin{figure}[!t]
\noindent\begin{picture}(246,350)
\put(-2,5){\psfig{figure=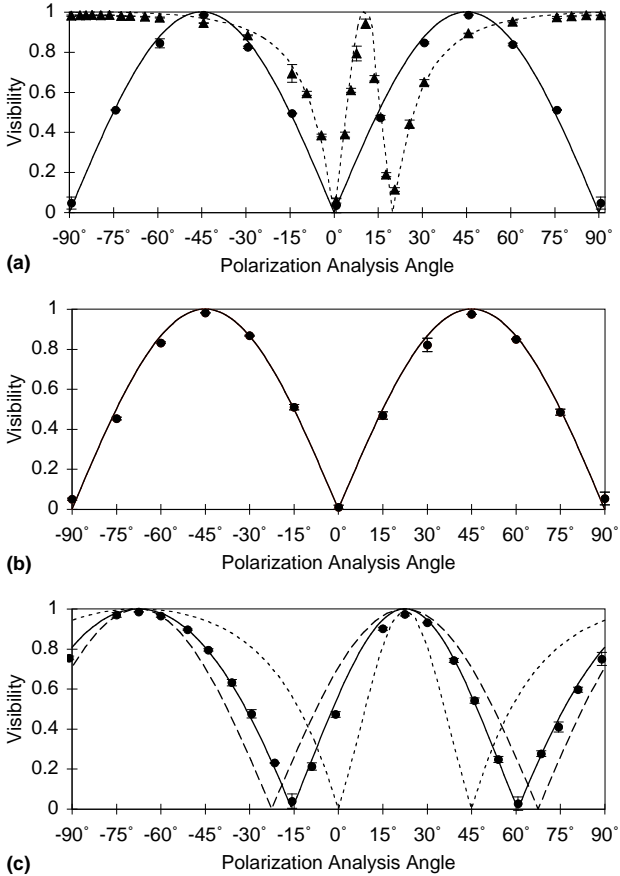,width=\columnwidth}}
\end{picture}
\caption[QE]{\label{fig:QE}
``Quantum eraser'' data and theory curves for various input states.
The minima on the curves correspond to analysis that transmits light
from only one or the other path; the maxima fall midway between these
minima.
(a) A purely vertically-polarized input ($\equiv 90^{\circ}$), with the
polarization rotated by the HWP in path 1 by $90^{\circ}$ (circles,
solid line; $\theta_{\rm{HWP}}=45^{\circ}$) or $20^{\circ}$
(triangles, dashed line; $\theta_{\rm{HWP}}=10^{\circ}$);
(b) a completely-mixed state, with $\theta_{\rm{HWP}}=45^{\circ}$;
and (c) a partially-mixed state (1:2 pure to mixed; circles,
solid line), with the HWP at $\theta_{\rm{HWP}}=22.5^{\circ}$ ---
the dotted and dashed curves show the corresponding theoretical
predictions for pure and completely-mixed states, respectively.}
\end{figure}

Figure \ref{fig:QE}a shows quantum eraser data under the condition that
a pure vertical photon is input to the interferometer, and rotated by
the HWP in path 1 by either $90^{\circ}$ or $20^{\circ}$.
The visibility on detector 1 {\em after\/} the analyzer can assume any  
value from $0$ to $1$, the latter case being a complete quantum erasure.
Even for a completely-mixed state,
it is still possible to recover interference (Fig.~\ref{fig:QE}b).
With no WW information to erase, this {\em non-erasing
quantum erasure\/} may seem quite remarkable at first.
However, the essential feature of quantum erasure is not that it
destroys the possibly available WW information, but that it sorts
the photons into sub-ensembles (depending on the quantum state of the 
WWM) each exhibiting high-visibility fringes.
Complete interference is recoverable by analyzing along the eigenmodes 
of the internal HWP --- along one axis we see fringes, along the other 
we see ``anti-fringes,'' shifted by $180^{\circ}$ \cite{ref:quartz}.
More generally, one post-selects one of the WWM eigenstates as 
determined by the interaction Hamiltonian of the interfering quantum 
system and the WWM \cite{ref:ZNatf}.

For a partially-mixed WWM state, just as the value of ${V^2 + K^2}$ lies
partway between $1$ (pure state) and $0$ (completely-mixed state), the
analysis angles yielding zero visibility also fall between those for 
pure and mixed states (Fig.~\ref{fig:QE}c).
Quantitatively, for a fractional purity $s$, the angles are at
$\theta_{\rm HWP} \pm 1/2\arccos\biglb(s\cos(2\theta_{\rm HWP})\bigrb)$;
consult Ref.\ \cite{ref:Garda} for further details.

A convenient geometrical visualization of our results can be had
by considering the polarization analysis in the Poincar\'{e}
sphere representation \cite{ref:Poincare}, in which all
linear polarizations lie on the equator of the sphere, circular
polarizations lie on the poles, and arbitrary elliptical states lie in
between.  
Any two orthogonal states lie diametrically opposed on the sphere.  
For pure polarization input states to our interferometer, there are in
general exactly two points on the sphere for which the interference
visibility will be exactly equal to zero. 
These correspond to the polarizations where a
detector sees light from only one of the interferometer paths.
Along the entire great circle that bisects
the line connecting these two points, the quantum eraser will yield
perfect visibility. 
Curiously, the situation for a completely-mixed input state is reversed.
Here there are in general exactly two polarization states for which the
quantum eraser recovers unit visibility, corresponding to the eigenmodes of
the polarization elements inside the interferometer; on the great circle
equidistant from these two points the Visibility vanishes.  
For example, in some mixed-state experiments described in Ref.\
\cite{ref:Garda}, the eigenmodes are the poles on the Poincar\'{e} sphere,
and the great circle corresponds to the equator  --- no visibility is
observed for any linear polarization analysis.

\section{Discussion}\label{sec:discuss}
Our results demonstrate the validity of Eq.\ (\ref{eq:WPDineq}) at the
percent level. 
Moreover, they highlight some features associated with
mixed states, which may not have been widely appreciated.  
Namely, that it is possible for both the interference visibility and the path
distinguishability to equal zero. 
We have also seen that in some cases where the visibility is intrinsically
equal to zero, it is possible to perform quantum erasure on the photons, and
recover the interference. 
Remarkably, this is true even when the input state is completely mixed, and
there exists no WW information to erase.  
The operation of the polarizer is essentially to {\em select\/} a
sub-ensemble of photons.  
Depending on how this selection is performed, we may recover fringes,
anti-fringes, no fringes, or any intermediate case.

The WW labeling in our experiment arose from an {\em entanglement\/}
between the photon's spatial mode and polarization state.
It could just as well have been with another photon altogether, as 
in the experiments in \cite{ref:QErasers1}, or even with a 
different kind of quantum system \cite{ref:Eichmann}.
The same results are predicted, as long as the WW information is stored
in a 2-state quantum system (e.g., internal energy states, polarization,
spin, etc.).
More generally, our findings are extendible to analogous experiments
with quanta of different kinds such as,
for example, interferometers with electrons \cite{ref:Buks},
neutrons \cite{ref:neutron1}, or atoms \cite{ref:Rempe1,ref:Rempe2}.

To counter a possible misunderstanding let us note that, quite generally,
entanglement concerns different degrees of freedom (DoF's), not different
particles.
For certain purposes, such as quantum dense coding \cite{ref:denscod} or
quantum teleportation \cite{ref:teleport}, it is essential that the entangled
DoF's be carried by different particles and can thus be manipulated at a
distance.
For other purposes, however, one can just as well entangle an internal DoF
of the interfering particle itself with its center-of-mass DoF
\cite{ref:Grover}.  
In our experiment the photon's polarization DoF is entangled with the
spatial mode DoF represented by the binary alternative ``reflected at the
entry beam splitter, or transmitted?'' 
Analogously, hyperfine levels of an atom were used to mark its path in the
experiments of Refs.\ \cite{ref:Rempe1,ref:Rempe2}.

In the extreme situation of perfect WW distinguishability, the entangled
state is of the form stated in Sec.~\ref{ssec:WPDpure}, namely
${|\psi\rangle \propto |1\rangle|{\sf H}\rangle
+ e^{i \phi}|2\rangle|{\sf V}\rangle}$.
Appropriate measurements on the spatial DoF (defined by $|1\rangle$ and
$|2\rangle$) and the polarization DoF ($|{\sf H}\rangle$ and 
$|{\sf V}\rangle$) would show that the entanglement is indeed so strong that
Bell's inequality \cite{ref:Bell} is violated --- clear evidence that a
description based solely on classical electrodynamics cannot account for all
features of our experiment.
Of course, inasmuch as one cannot satisfy the implicit assumption that the
measurements on the entangled subsystems be space-like separated, this
violation of Bell's inequality implies nothing about the success or failure
of local-hidden-variable theories; however, this is not relevant here.

Finally we'd like to mention that further progress was made since the
completion of the work reported here.
Experimental tests of more sophisticated inequalities than (\ref{eq:WPDineq})
were performed \cite{ref:DR99}, and there was progress in theory as well
\cite{ref:EB99}.
In particular, the quantitative aspects of quantum erasure were investigated
beyond the initial stage reached in Ref.\ \cite{ref:BjKa}.

\section*{Acknowledgments}
BGE is grateful to Helmut Rauch and collaborators for their hospitality
at the Atominstitut in Vienna, where part of this work was done,
and he thanks the Technical University of Vienna for financial support.
PGK and PDDS acknowledge Andrew White for helpful discussions and
assistance. Correspondence should be addressed to PGK 
(email: Kwiat@lanl.gov).


\begin{references}
\bibitem{ref:compressed}
The angle of incidence on the beam splitter was set to $10^{\circ}$, in
order to minimize polarization variations in the reflection and
transmission amplitudes -- the resulting beamsplitter reflectivies and
transmittivities were found to lie in the range $0.49$ to $0.51$ for
all polarizations.
\bibitem{ref:optimal}
For linear polarization states, the optimal Knowledge-measurement axes
lie exactly between the axes which would equalize the amplitudes from
the two paths (and for which the Visibility is maximum), i.e., if the
light coming from paths 1 and 2 is polarized at $\phi_1$ and $\phi_2$,
then the optimal Knowledge basis is at
$(\phi_1+\phi_2)/2 \pm 45^{\circ}$.

\enlargethispage{-29.5\baselineskip}
\end{references}
\end{document}